# Minority Spin Condensate in Spin-Polarized Superfluid $^3$He A$_1$ Phase

A. Yamaguchi[1], S. Kobayashi[1], H. Ishimoto[1] and H. Kojima[2]

[1]*Institute for Solid State Physics, University of Tokyo, Kashiwa, Chiba 277-8581 Japan*

[2]*Serin Physics Laboratory, Rutgers University, Piscataway NJ 08854 USA*

**The fascinating magnetic properties of $^3$He in its various phases originate from the interactions among nuclear spins in $^3$He.[1]  The spin-polarized "ferromagnetic" superfluid $^3$He A$_1$ phase,[2] which forms below 3 mK in external magnetic field, serves as a material in which theoretical notions of fundamental magnetic processes and macroscopic quantum spin phenomena may be tested. Conventionally, the superfluid component of the A$_1$ phase is understood to contain only the majority spin condensate having energetically-favoured paired spins directed along the external field and no minority spin condensate having paired spins in the opposite direction.[3-6]  Owing to difficulties in satisfying both ultra low temperature and high magnetic field to produce a substantial phase space, little study of spin dynamics phenomena exists to test the conventional view of the A$_1$ phase.  Here, we report on the development of a novel mechanical spin density detector while meeting both requirements and on the first measurements of the spin relaxation in the A$_1$ phase as functions of temperature, pressure and magnetic field.  The mechanical spin detector is based in principle on the unique magnetic fountain effect[7] occurring only in the A$_1$ phase (delineated between two transition temperatures, T$_{c1}$ and T$_{c2}$).    In the high temperature range near T$_{c1}$, the measured spin relaxation time is long, as expected.[2,8,9]  Unexpectedly, the spin relaxation rate increases rapidly as the temperature is decreased towards T$_{c2}$.  Our measurements, together with Leggett-Takagi theory,[5] demonstrate that a minute presence of minority spin pairs is responsible for the unexpected spin relaxation**

**phenomena in A₁ phase. Thus, the long-held conventional view[2] of A₁ phase**

**containing purely majority spin condensate is inadequate. In our device, the spin-**

**polarized superfluid motion can be induced both magnetically and mechanically.**

**Our work demonstrates for the first time feasibility of increasing spin polarization**

**by a mechanical spin filtering process.**

The dipolar-interaction between nuclear spins in bulk normal liquid $^3$He results in an intrinsic longitudinal spin relaxation time[9] which diverges at low temperature($T$) as $T^{-2}$.

Magnetic relaxation of liquid $^3$He in any experiment at low temperatures, however,

inevitably involves both the intrinsic and an additional, usually much more

rapid, extrinsic relaxation occurring at the surfaces of the apparatus or particles

introduced into it.[10] The magnetic fluctuations in the dense solid-like $^3$He layer adjacent

to surfaces are thought to be responsible for the magnetic relaxation time linearly

proportional to $T$ and to magnetic field H.[11,12] Below the superfluid transition

temperatures, new mechanisms of intrinsic magnetic relaxation mediated by the

condensate spin pairs become possible.[4,5] Measurements[13] of NMR linewidths in the

$^3$He A phase in small magnetic fields have been interpreted in terms of such intrinsic

relaxation mechanisms. We present the first measurements and their analysis of the

intrinsic spin relaxation using an unusual mechanical spin density detector in the bulk

$^3$He A₁ phase in high magnetic field.

Consider a small detector chamber connected to a much larger reservoir chamber filled with $^3$He A₁ phase via a superleak (see Fig. 1 and Supplementary Information). A rapid increase (or decrease) in spin polarization in the small chamber is generated by the superflow induced by an applied magnetic field gradient across the superleak. The superflow in the superleak (spin filter) carries the spin-polarized superfluid component of A₁ phase. If the superleak is ideal, the spin polarization gradient across the superleak, in the absence of spin relaxation, would be balanced by a steady magnetic fountain pressure difference between the two chambers.[7]  In the presence of spin relaxation, however, the pressure difference relaxes, and the observed fountain pressure relaxation gives a direct measure of the spin polarization relaxation.

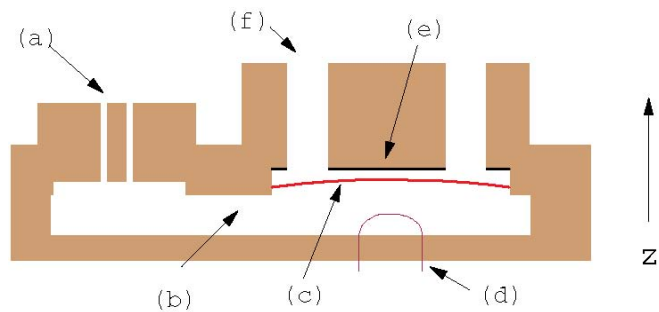

Fig. 1 Schematic cross sectional view of spin density detector.  (a) superleak channels (two parallel channels made by etching out 18 μm thick, 3.0 mm long, and 3.6 mm wide aluminium foils cast into Stycast 1266), (b) detector chamber of volume 0.3 cm$^3$, (c) lightly stretched circular flexible membrane with a tension $1.9 \times 10^5$ dyne/cm (9 mm diameter and 9 μm thick aluminium coated Mylar membrane), (d) vibrating wire viscometer, (e) stationary metal film electrode, (f) four vent holes (1.5 mm diameter and 4 mm long, only two are shown).   The working principle is based on the magnetic fountain effect in $^3$He $A_1$ phase.  The whole detector assembly is immersed in a large reservoir of liquid $^3$He in a uniform static magnetic field.  The static and gradient magnetic fields are applied along $-\mathbf{z}$ (down) direction parallel to the superleak channels.  The displacement of the flexible diaphragm is detected by measuring the changes in the capacitance between the fixed electrode and the diaphragm.  The displacement gives a measure of the induced (differential) magnetic fountain pressure across the superleak.  The diaphragm is located on the same side of the detector chamber and the vent holes are provided such that the measured differential pressure reflects more accurately that across the superleak.  All structural parts except those noted are fabricated with Stycast 1266 epoxy.  Electrical leads are not shown.

The principal mechanism that governs the quasistatic (superfluid acceleration is ignored) response of our detector is that the superfluid maintains equal chemical potential at both ends of the superleak. This leads to the magnetic fountain effect[7] expressed by: $\delta P/\rho + (\gamma \hbar /2m)(\gamma \delta S/\chi - \delta H) = 0$, where $\delta P$, $\delta S$ and $\delta H$ are the differential pressure, spin density and magnetic field across the superleak, respectively, m is the mass of $^{3}$He atom, $\rho$ the mass density, $\gamma$ the gyromagnetic ratio, and $\chi$ the magnetic susceptibility[1]. The induced differential pressure $\delta P$ is related to the displacement $Z$ of the flexible membrane (of area $A_m$ and tension $\sigma$) by $8\pi\sigma Z = A_m\delta P$. The membrane displacement is a direct measure of the magnetically induced pressure. Since the superleak is imperfect, the induced pressure produces small concurrent normal fluid flows. The measured membrane dynamics is determined by both the relaxation (with relaxation time $T_1$) in spin density and the normal fluid flow (with relaxation time $\tau_n$). Combining these, the conservation of mass and the spin relaxation leads to the membrane relaxation time $\tau$:

$$\frac{1}{\tau} = \left(\frac{1}{\tau_n} + \frac{\alpha}{T_1}\right)\left(\frac{1}{1+\alpha}\right), \quad\quad\quad (1)$$

where $\alpha = 8\pi\sigma\chi V/(\hbar\gamma\rho A_m/2m)^2 (= 24)$ is the mechanical to magnetic energy density ratio and $V(= 0.30 \text{ cm}^3)$ is the volume of the detector(see Supplementary Information).

Examples of the membrane displacement are shown in Fig. 2 when a magnetic field difference $\delta H \sim 0.63$ mT is applied across the superleak within a typical time of 100 ms. As expected, there is no measurable response in the normal(T > 2.50 mK) and $A_2$ (T < 2.11 mK) phases. Within the $A_1$ phase, the displacement increases linearly in time to a maximum, which depends on the applied field difference and the field ramp rate. For a large enough field difference, $dZ/dt$ tends to saturate presumably limited by a critical velocity (measured to be ~ 0.5 mm/s, consistent with others[14]) in the superleak. Subsequent to reaching the maximum, the displacement decays exponentially in time at all temperatures except near $T_{c1}$. The measured relaxation time does not depend on the maximum displacement amplitude.

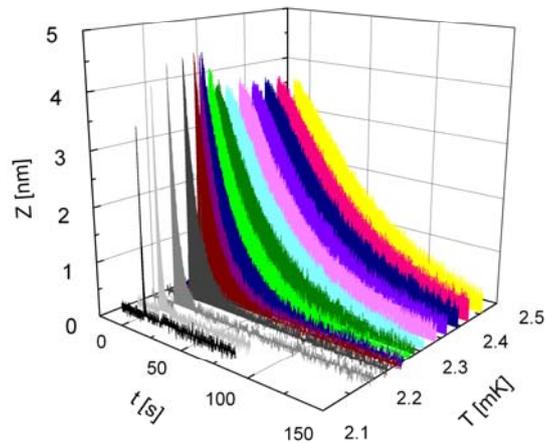

Fig. 2 Time response of detector membrane displacement to applied step in field gradient at t = 0. The liquid pressure is 21 bar and the applied static field is 8 T. The colours distinguish 14 temperatures (see Supplementary Information) at which data are shown. Many more data are taken than shown. There is no response outside of $A_1$ phase temperature range. The onset temperatures of the magnetic fountain effect coincide with the kinks in shear viscosity[15] at $T_{c1}$ and $T_{c2}$ as detected by the viscometer placed in the detector chamber. After the initial rise, the response decays exponentially.

The decaying portion of the measured response is fitted with an exponential function with time constant $\tau$. Fig. 3 shows $\tau$ as a function of normalized reduced temperature, $r = (T_{c1} - T)/(T_{c1} - T_{c2})$, where $T_{c1} - T_{c2} = 0.052H$ mK/T, at a pressure of 21 bar in static magnetic fields up to 8 T. Non-exponential and erratic relaxations near $T_{c1}$ where $r < 0.18$ are excluded from Fig. 3. Clearly, the observed relaxation time is distinct from that in normal fluid. At a given field, $\tau$ gradually decreases as the temperature decreases from $r = 0$ and it rapidly decreases over a relatively small temperature interval at a "kink" temperature(indicated by arrow). The kink temperature tends toward $T_{c1}$ as the static field is decreased towards zero. The relaxation time finally tends towards zero as the temperature approaches $T_{c2}$ ($r = 1$).

At a given temperature $r$, $\tau$ increases as the static field is increased. The presence of the "kinks" introduces more complication to the field dependence at low fields and small values of $r$. However, it is clear that $\tau$ tends to increase with increasing applied static field at a given value of $r$. Except in the region of kinks, $\tau$ increases linearly with applied field above 1 T (see supplementary information). This field dependence is reminiscent of the spin relaxation mediated by the solid-like $^3$He layer at the boundaries observed in magnetic relaxation in liquid $^3$He immersed in fluorocarbon particles,[11] but this is not the case as discussed below.

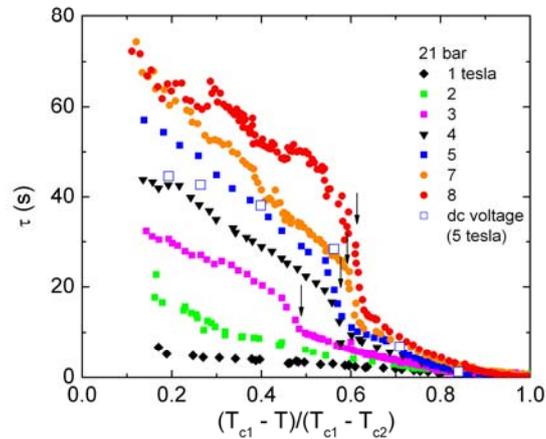

Fig. 3 Measured relaxation time versus normalized reduced temperature at 21 bar. Relaxation time tends to vanish as $T_{c2}$ is approached, increases as $T_{c1}$ is approached, and shows a sharp change at an intermediate temperature which depends on the applied static field. An example of data for the measured relaxation subsequent to induced spin polarization by an applied dc voltage to the membrane (see text) is shown (open squares) for the static field of 5 T.

Instead of using a magnetic field gradient, spin polarization may be induced by mechanical pumping by applying a step voltage to the flexible membrane with respect to the fixed electrode thereby drawing in spin-polarized superfluid into the chamber. The process is in effect spin pumping through the superleak spin filter. Studies of time response to such mechanical manipulation of the spin-polarized superfluid show that the observed relaxation time is the same as those observed subsequent to magnetically induced spin flow under the same conditions as in Fig. 3. This observation proves that the mechanically driven superflow induces spin polarization and that the relaxation time is not dependent on the presence of applied magnetic field gradient.

An important check to distinguish the source of the observed spin relaxation as the bulk liquid or the solid-like boundary layer at the walls is to replace the $^3$He boundary layer by magnetically inert $^4$He, which preferentially covers the boundary walls.[10] In a separate experiment, a sufficient amount of $^4$He was introduced to cover all surfaces with five monolayers of $^4$He, which should remove the magnetically active layer.(See Freeman et al.,[16] for example). The observed relaxation times with the $^4$He coverage are the same within 15 % as those in pure $^3$He at all temperatures and magnetic fields. This provides strong evidence that the observed spin relaxation occurs within the bulk liquid.

Surprisingly, the measured relaxation time tends to vanish as $T_{c2}$ is approached at all magnetic fields. To highlight the region near $T_{c2}$, and to make direct comparison with theory, $T_1^{-1}$ is extracted from the measured the relaxation rate $\tau^{-1}$ using Eq. (1). The result is plotted against $t = (T - T_{c2})/T_{c2}$ in Fig. 4 for the data at 2 T. To our knowledge, no spin transport mechanism has been predicted to show such an increase in spin relaxation within the $A_1$ phase near $T_{c2}$. Below, we interpret the behaviour near $T_{c2}$ with the theory by Leggett and Takagi[5](LT) and the predicted presence of a tiny amount of minority spin condensate(MSC) in the $A_1$ phase by Monien and Tewordt[17,18](MT).

LT[5] introduced a two fluid model of spin relaxation in superfluid $^3$He. Very briefly, the total spin is decomposed into those spins carried by the normal and superfluid components. When the spin polarization of the superfluid component changes, its associated magnetic field (or chemical potential) can change essentially instantaneously (within a time scale of $\hbar/\Delta$) with its magnetic susceptibility at constant normal component spin. This induces a chemical potential difference between the superfluid and normal fluid components. The spins in the two components approach equilibrium by spin-conserving collisions with a characteristic relaxation time. During these two processes, the spin polarization and field can get out of phase and produce dissipation and an associated spin relaxation rate.

According to the above LT theory, the relaxation rate of longitudinal magnetization in the A phase is given by

$$1/T_1 = Y_2 \tau_n \chi \, \Omega_\parallel^2/(1-Y_2)\chi_o \qquad (2)$$

where $Y_2$ is the Yoshida function of the second kind[19], $\chi/\chi_o = (1 + Z_o/4)^{-1}$, $\chi_0$ the magnetic susceptibility in the absence of Fermi liquid effects[1], $Z_o$ a Landau parameter[1], $\tau_n$ the quasiparticle relaxation time at $T_c$, and $\Omega_\parallel$ the longitudinal magnetic resonance frequency. Comparison of Eq. (2) with experiment has been difficult owing to suspected presence of spin currents. Qualitative agreement, however, has been seen in the measurements of longitudinal resonance line width[13] and of the longitudinal relaxation time[20]. LT conjectured that the mechanism described above would be applicable to the $A_1$ phase if the changes in magnetic fields and polarizations are small compared to their original values.

An applied magnetic field $H$ with associated field energy $\eta'H$ tends to create $A_1$ phase with ↑↑ spin pairs with an associated energy gap $\Delta_{\uparrow\uparrow}$ in accordance with particle-hole asymmetry. This is the "conventional" description of $A_1$ phase and there is no longitudinal magnetic resonance, $\Omega_\parallel = 0$. Eq. (2) then implies $T_1$ would diverge, in disagreement with experiment. However, MT[17] show that inclusion of dipolar interaction with energy scale[6] $g_D$ tends to create MSC with opposite ↓↓ spin pairs with energy gap $\Delta_{\downarrow\downarrow}$. MT find that $\Delta_{\downarrow\downarrow}/\Delta_{\uparrow\uparrow} \sim g_D/\eta'H \sim 10^{-4}/H[T]$. The presence of this minute MSC population implies a finite longitudinal nuclear magnetic resonance frequency, contrary to the "conventional" description of the $A_1$ phase. We apply this new $\Omega_\parallel$ to Eq. (2). The miniscule MSC population turns out to produce a huge influence on spin relaxation!

To apply the LT theory to the $A_1$ phase, the function $Y_2$ is transformed to $Y_2^{A1}$ as follows: In the tabulation[19] of $Y_2$, we set $Y_2(T/T_c = 1) = Y_2^{A1}(r = 1)$ and $Y_2(T/T_c = (T_{c1}-T_{c2})/T_{c1}) = Y_2^{A1}(r = 0)$ for given H and interpolating in between by setting $Y_2(T/T_c) = Y_2^{A1}(r)$. The parallel resonance frequency $\Omega_\parallel$ in the $A_1$ phase in 2 T is taken from Fig. 9 of MT.[17] The evaluated relaxation rate from Eq. (2) with <u>no adjusted parameter</u> is shown in Fig. 4. The measured and theoretical relaxation rates, where $t < 0.01$, are within a factor of four. The temperature dependence of the theoretical $T_1^{-1}$ is quite similar to that of the data. The theoretical relaxation rate critically depends on the calculated longitudinal resonance frequency in the $A_1$ phase. In view of the unknown influence on the parallel resonance frequency by texture and detector chamber geometry, the agreement between theory and experiment may be considered good. Comparison between the measured dependences of $T_1^{-1}$ on magnetic field and pressure to those expected from the theory show equally good agreement (see supplementary information).

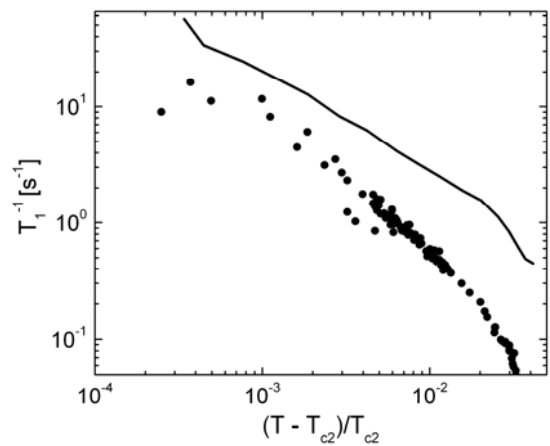

Fig. 4 Dependence of longitudinal relaxation rate on reduced temperature relative to $T_{c2}$. The dots are those evaluated from the measured relaxation time using Eq. (1) and the line is $T_1^{-1}$ based on the presence

of minority spin condensate[17] and Leggett and Takagi theory[5] on spin relaxation without any parameter adjustment(see text).

Osheroff and Anderson[21] measured NMR in both the $A_1$ and $A_2$ phases in the vicinity of $T_{c2}$ in 1.5 T. They concluded that no longitudinal resonance occurred in the $A_1$. The predicted[17] longitudinal frequency of 2.5 KHz in the $A_1$ phase was unlikely to be detectable in their experiment. On the other hand, an increasing longitudinal relaxation rate $1/T_1$ was seen, in agreement with our observations, in the $A_1$ phase as $T_{c2}$ was approached in a field of 0.305 T.[20] It has been observed that the spin-entropy (second sound) wave propagation[22] suffers an extra "anomalous attenuation" in the vicinity of $T_{c2}$. The presence of MSC is likely the origin of the extra attenuation.

Our conclusion that the $A_1$ phase contains an MSC implies that it is an incipient $A_2$ phase, which would not support magnetic fountain effect nor spin-entropy wave. Why then does the presence of MSC increase the spin relaxation rate but does it not suppress the magnetic fountain effect? To address this question, we estimate the pair breaking critical velocity[2] for the MSC, $v_c \sim \Delta_{\downarrow\downarrow}/p_F$, where $p_F$ is the Fermi momentum[1] and $\Delta_{\downarrow\downarrow}$ is estimated according to MT(see above). At 21 bar and 5 tesla, we find $v_c$ is at most of order 1 μm/s. The superflow velocity (greater than 50 μm/s) in our superleak channels during ramping up of the applied magnetic field gradient certainly exceeds $v_c$. When the MSC is broken by the flow, the conventional $A_1$ phase is restored along with the magnetic fountain effect. The pair breaking velocity increases as the magnetic field is decreased and this might be contributing in part to the increasing deviation between the LT mechanism theory and the relaxation time data at low magnetic fields (see Fig. 3 of Supplementary Information). The small MSC pair breaking velocity was very likely exceeded during the oscillatory period of superflow (typically 10 μm/s) in spin-entropy wave experiments.[22] All previous spin-entropy wave observations have been made at frequencies greater than $10^2$ Hz, which exceeds the spin relaxation rate of at most 20 s$^{-1}$ observed here.  Thus, once the system becomes supercritical, it tends to remain
essentially in the conventional $A_1$ phase state allowing spin-entropy wave propagation
in the "high" frequency measurements.  Magnetic fountain effect experiments have
uniquely uncovered novel phenomena previously undetected in the spin-polarized
superfluid $^3$He.

**Acknowledgements.**

We thank Ryuichi Masutomi and Kota Kimura for their contributions in the early stages of the experiments and Tony Leggett for correspondence. This work was supported in part by Grant-in-Aid for Scientific Research on Priority Areas (Grant No. 17071004) of MEXT Japan and JPSJ, and by the Condensed Matter Physics and East Asia and Pacific Programs of the National Science Foundation of U.S.A.

Correspondence and requests for materials should be addressed to Harry Kojima (kojima@physics.rutgers.edu).